\documentclass{article}

     \usepackage[final]{neurips_2019}

\usepackage[utf8]{inputenc} 
\usepackage[T1]{fontenc}    
\usepackage{hyperref}       
\usepackage{cleveref}
\usepackage{url}            
\usepackage{booktabs}       
\usepackage{amsfonts}       
\usepackage{nicefrac}       
\usepackage{microtype}      
\usepackage{multirow}
\usepackage{pgfplots}

\crefformat{section}{\S#2#1#3} 
\crefformat{subsection}{\S#2#1#3}
\crefformat{subsubsection}{\S#2#1#3}

\title{Generative Image Translation for Data Augmentation in Colorectal Histopathology Images}

\author{
  Jerry Wei$^1$, Arief Suriawinata$^2$, Louis Vaickus$^2$, Bing Ren$^2$, \\ \vspace{4mm}
  \textbf{Xiaoying Liu$^2$, Jason Wei$^{1}$, Saeed Hassanpour$^{1*}$}\\ 
  $^1$Dartmouth College, Hanover, NH, USA\\
  $^2$Dartmouth-Hitchcock Medical Center, Lebanon, NH, USA\\
  $^* $\texttt{saeed.hassanpour@dartmouth.edu}\\
}

\begin{document}

\maketitle

\begin{abstract}
  We present an image translation approach to generate augmented data for mitigating data imbalances in a dataset of histopathology images of colorectal polyps, adenomatous tumors that can lead to colorectal cancer if left untreated.
  By applying cycle-consistent generative adversarial networks (CycleGANs) to a source domain of normal colonic mucosa images, we generate synthetic colorectal polyp images that belong to diagnostically less common polyp classes. 
  Generated images maintain the general structure of their source image but exhibit adenomatous features that can be enhanced with our proposed filtration module, called Path-Rank-Filter. 
  We evaluate the quality of generated images through Turing tests with four gastrointestinal pathologists, finding that at least two of the four pathologists could not identify generated images at a statistically significant level.
  Finally, we demonstrate that using CycleGAN-generated images to augment training data improves the AUC of a convolutional neural network for detecting sessile serrated adenomas by over 10\%, suggesting that our approach might warrant further research for other histopathology image classification tasks. 
\end{abstract}

\section{Introduction}

Accurately analyzing medical images with deep learning classifiers often requires large, balanced datasets. 
For many diseases, however, the distribution of disease sub-classes in collected datasets is heavily skewed by each class's prevalence among patients, and so detecting rare diseases in medical images with deep learning can be challenging.
In these situations, a reliable method of data augmentation can mitigate the effects of data imbalance by preventing overfitting and thus improving overall performance. 

Previous work in data augmentation includes both traditional augmentation methods (rotations, flips, color jittering, etc.) and, more recently, generative models that synthesize completely new images. 
Since their development, generative adversarial networks (GANs) \citep{Goodfellow2014}, which use noise as an input variable, have been a popular method of generating augmented data for improving image classification \citep{DBLP:journals/corr/abs-1712-04621, DBLP:journals/corr/abs-1712-01636}.
We hypothesized that, in the field of medical image analysis, data from one class might contain useful information to synthesize new data for another. 
As such, generative image translation models might suit this task better than models that do not account for information in other classes (e.g., models that use random noise as a basis for image generation).


In this paper, we present an image translation model for generating synthetic colorectal histopathology images. 
Since adenomatous preneoplastic polyps always originate from normal colonic mucosa, we use normal colonic mucosa as a source domain to generate synthetic images that are similar in structure but present adenomatous features. Our work makes the following contributions:
\begin{enumerate}
    \item We demonstrate an image translation model that generates synthetic images of adenomatous colorectal polyps and propose a filtration module called Path-Rank-Filter that enhances the presence of adenomatous features in generated images. 
    \item We evaluate the quality of generated images through Turing tests with four gastrointestinal pathologists, finding that for the two adenomatous polyp classes tested, at least two of four pathologists could not distinguish between synthetic and real polyp images at a statistically significant level. 
    \item We show that using generated images as augmented data for training improves the AUC of a convolutional neural network in detecting sessile serrated adenomas by over 10\%, indicating that our approach might be useful for other histopathology image classification tasks. 
\end{enumerate}
Our code for this project is   \href{https://github.com/BMIRDS/HistoGAN}{publicly available.}\footnote{https://github.com/BMIRDS/HistoGAN}

\section{Related Work}
\label{gen_inst}


Generative adversarial networks (GANs) have commonly been used in the field of medical image analysis. 
For magnetic resonance imaging (MRI) scans, \citet{Nie2016} used context-aware GANs to generate computed topography (CT) images from MRIs, and \citet{Yang2018} used conditional GANs (cGANs) to generate target modality MRIs given a particular source modality MRI. 
Furthermore, \citet{Dar2018} used cGANs to generate fake T1 and T2 MRIs and used an improved methodology by using end-to-end training of GANs that synthesize target images given source images.
\citet{Hiasa2018} also translated MRIs to CT images with CycleGANs, adding a gradient consistency loss to encourage edge alignment between images. 
\citet{Salehinejad2018} used DCGANs to generate fake chest x-ray images from real ones, though the resulting fake images were at a lower resolution than real images, and \citet{Wang2018} used cGANs to reduce artifacts in CT images by learning to map an artifact-affected CT image to an artifact-free CT image. 

In the field of histopathology in particular, many studies have used GANs for both image generation and image translation. 
Both \citet{Bayramoglu2018} and \citet{Rana2019} used cGANs to virtually stain Haemotoxylin and Eosin (H$\&$E) lung tissue histopathology.
Similarly, \citet{Hou2017} and \citet{Quiros2019} generated fake histopathology samples with GANs, and \citet{Burlingame2018} used cGANs to translate pancreas tumors from H$\&$E-stained to immunofluorescent. 
In terms of stain normalization, \citet{Bentaieb2017} used a GAN to normalize tissue samples in order to remove natural discolorations from tissue staining, and \citet{Cho2017} performed stain style transfer by replacing stain normalization models with cGANs.
Moreover, \citet{Zanjani2018} integrated a Convolutional Neural Network (CNN) and Gaussian Mixture Model to jointly optimize the modeling and normalizing of color and intensity in H$\&$E stained images.

In terms of data augmentation, both conventional methods and GANs have been used in previous research. 
\citet{Hussain2017} found that effective methods of data augmentation for images primarily include strategies such as flips, Gaussian noise, jittering, Gaussian blurring, and rotations, and \citet{Li2010} addressed class imbalances by oversampling abnormal classes and undersampling normal classes.
For generative methods, \citet{Bass2019} synthesized augmented biomedical images with convolutional capsule GANs. 
Additionally, \citet{Gupta2019} used CycleGANs on x-ray images to generate augmented images of bone lesions, which were then added to a training set to improve a bone lesion classifier's AUC by 5$\%$.
Both papers, however, did not manually evaluate the quality of their generated images, and \citet{Gupta2019} did not have extensive ablation studies to provide insight on how their method could be applied to other datasets. 

In our study, we apply CycleGAN to a colorectal histopathology image dataset to generate augmented data.
We propose a filtration module called Path-Rank-Filter that improves the quality of generated images for some classes and perform extensive ablation studies. 
Furthermore, we evaluate our generated images manually with four pathologists and compare our CycleGAN model's ability to improve classifier performance with that of two other generative models: DCGAN \citep{Radford2015} and DiscoGAN \citep{Kim2017}.

\section{Image Translation in Colorectal Histopathology Images}

Here, we discuss our approach for applying generative image translation to a dataset of colorectal histopathology images. 
We focus on cycle-consistent generative adversarial networks \citep{Zhu2017} and propose a simple filtration module called Path-Rank-Filter that enhances the adenomatous features in generated images.
Additionally, we describe the process of collecting our dataset as well as our experimental setup. 

\subsection{Cycle-Consistent Generative Adversarial Networks}


We use a cycle-consistent generative adversarial network (CycleGAN) \citep{Zhu2017} model to translate images of normal colonic mucosa to images of adenomatous colorectal polyps. 
Given two domains, $X$ and $Y$, with training samples $\{x_i\}^N_{i=1}$, where $x_i \in X$, and $\{y_i\}^N_{i=1}$, where $y_i \in Y$, CycleGAN learns the mapping $G: X \rightarrow Y$ for unpaired image translation. 
For colorectal polyp images, we set $X$ as normal colonic mucosa, which has many images, and $Y$ as a less common polyp type with few images (e.g., tubular adenoma or sessile serrated adenoma) so that we can mitigate the imbalance of class $Y$ by generating a set of augmented data $\{G(x_i)\}_{i=1}^{N}$ that presents features of domain $Y$. 
\subsection{Path-Rank-Filter}
\begin{figure}[ht]
    \setlength{\abovecaptionskip}{5pt}
    \includegraphics[width=\linewidth]{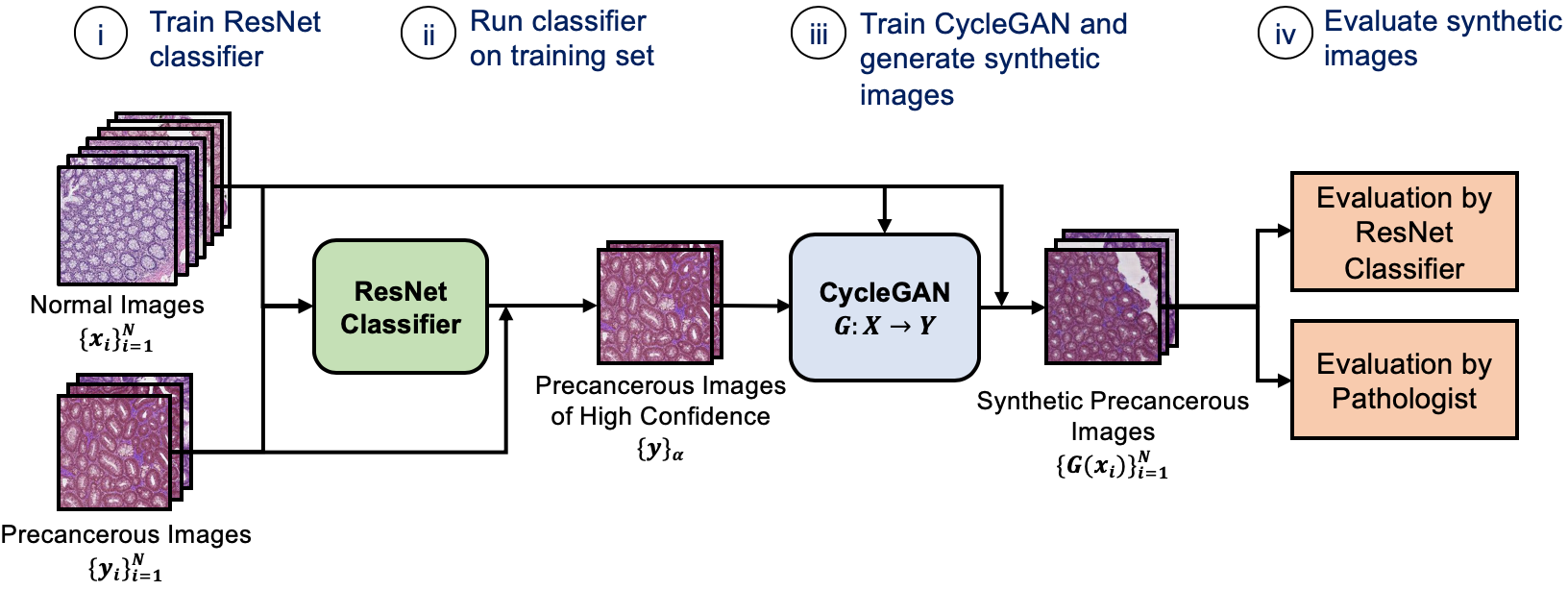}
    \caption{Process for generating synthetic histopathology images of rare colorectal polyp classes. Path-Rank-Filter (i-ii) enhances the adenomatous features in generated images by filtering the training data for CycleGAN for only images with strong adenomatous features.}
    \label{fig:filtration}
\end{figure}

Because histopathology images differ in nature from images in standard computer vision datasets (e.g., MNIST or ImageNet), we propose a module called Path-Rank-Filter that improves CycleGAN's performance specifically for histopathology images. 
Whereas distinguishing between common classes in computer vision (e.g., cats and dogs) is relatively straightforward, histopathology images can contain a range of histologic features that determine whether an image can be classified as adenomatous.
For instance, both an image with small amounts of tubular architectures and an image covered by tubular architectures would be classified by a pathologist as a tubular adenoma. 
We hypothesize that images with more prominent features will be more useful for training, and so instead of training a CycleGAN on the original $\{x_i\}^N_{i=1}$ and $\{y_i\}^N_{i=1}\ $, we introduce the following filtration process (Figure \ref{fig:filtration}):
\begin{enumerate}
    \item We train a ResNet \citep{He2015} $f$ to classify $X$ and $Y$. We define $f_Y(y_i)$ as the output probability of the ResNet for class $Y$ when given image $y_i$ as the input. 
    \item Then, we run the ResNet on all $\{y_i\}^N_{i=1}\ $. For some $\alpha \in (0, 1]$, we find $\{y\}_{\alpha} \subset \{y_i\}_{i=1}^{N}$ such that for all $y_i \in \{y\}_{\alpha}\ $, $f_Y(y_i)$ is in the highest $\alpha$ of all output probabilities $\{f_Y(y_i)\}^N_{i=1}\ $. 
    \item We train CycleGAN on $\{x_i\}^N_{i=1}$ and $\{y\}_{\alpha}$ instead of $\{x_i\}^N_{i=1}$ and  $\{y_i\}^N_{i=1}$.
\end{enumerate}

Path-Rank-Filter uses the knowledge that an adenomatous class $Y$ includes images with a range of histologic features. 
It thus finds the images with the strongest features that are most representative of class $Y$ and uses those images to train CycleGAN.


\subsection{Dataset Collection}

Our dataset of colorectal polyp images was collected from the Dartmouth-Hitchcock Medical Center in New Hampshire, USA, our tertiary medical institution.
We collected 427 high-resolution whole-slide images, which we split into a training set of 326 whole-slide images and a testing set of 101 whole-slide images.
For the training set, pathologists annotated all whole-slide images with bounding boxes representing regions of interest, for a total of 3517 variable-size image crops. 
Each image crop was labeled with a single class for the polyp type, which was either benign (normal or hyperplastic), or adenomatous (tubular adenoma, tubulovillous/villous adenoma, or sessile-serrated adenoma). 
The distributions of different classes in our training set is shown in Figure 2.

For the testing set, pathologists annotated the whole-slide images for fixed-size tiles of classic examples of polyp types ($224 \times 224$ pixels), and polyp type labels were verified by two pathologists so that our evaluation was as close to ground truth as possible. 
Our final testing set, which is used in \cref{auc}, had 261 hyperplastic polyp images and 39 sessile serrated adenoma images.

\begin{figure}[ht]
    \setlength{\abovecaptionskip}{5pt}
    \includegraphics[width=\linewidth]{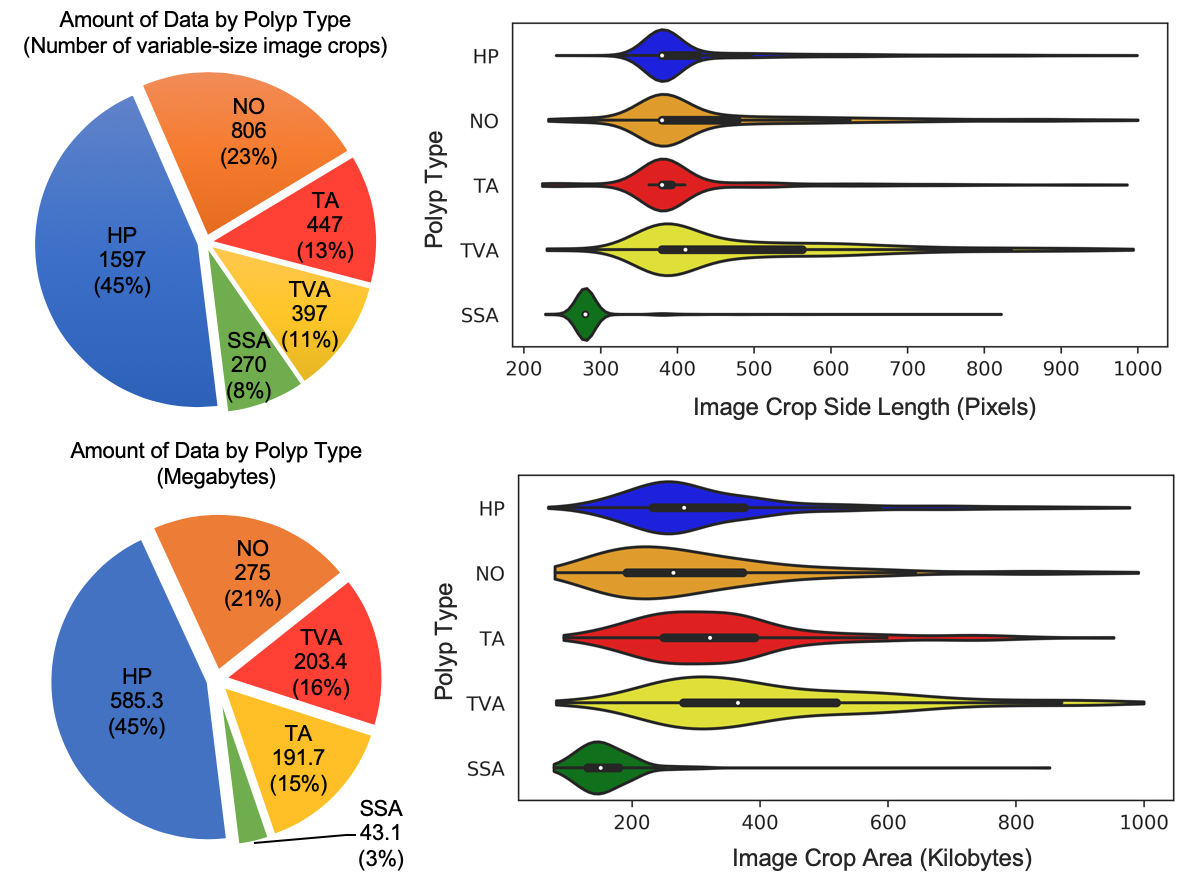}
    \caption{Distribution for collected dataset of colorectal polyp histopathology images. HP: hyperplastic polyp, NO: normal colonic mucosa, TVA: tubulovillous/villous adenoma, TA: tubular adenoma, SSA: sessile serrated adenoma. Two diagnostically relevant classes of adenomatous polyps, tubular adenoma (TA) and sessile serrated adenoma (SSA), comprise only 14.8\% and 3.3\% of the dataset, respectively. }
    \label{fig:dataset}
\end{figure}





\subsection{Experimental Setup and Motivation}

In this study, we set tubular adenoma (TA) and sessile serrated adenoma (SSA), two adenomatous polyp types that respectively account for only 14.8\% and 3.3\% of our dataset by size, as the target domains for data generation. 
As a source domain, we use normal colonic mucosa images, since both tubular and sessile serrated adenomas emerge as a result of cytological transformations on normal colonic mucosa. 

For all classifiers, we use the ResNet architecture \citep{He2015} and train each classifier for 20 epochs.  
We conducted an ablation study for our particular classification task and found that increasing the depth of the neural network did not substantially improve performance (Supplementary Figure \ref{fig:resnetlayers}).  
Thus, for all experiments, we used the model with the lowest number of parameters, ResNet-18, so that experiments can be replicated more quickly.

\section{Experiments}

We perform extensive experiments to evaluate the ability and usefulness of generative image translation on colorectal polyp histopathology images.
We measure the strength of our filtering method using a pre-trained classifier, finding that CycleGAN with Path-Rank-Filter generates images that are substantially closer to the target domain (i.e. exhibit more adenomatous features) than when Path-Rank-Filter is not used.
Next, we perform a clinical evaluation of our images by conducting a Turing test with four gastrointestinal pathologists, finding that three of the four pathologists could not differentiate at least half of the synthetic images from real images.
Finally, we evaluate how adding the generated images as augmented data for training a ResNet classifier can improve performance for detecting sessile serrated adenomas, a clinically important distinction in colorectal cancer screening.

While we limit the scope of this paper to a single source domain, normal colonic mucosa, we show qualitative results of experiments on other source domains in Supplementary Figure \ref{fig:transformationbase}.

\subsection{Enhancing Adenomatous Features with Path-Rank-Filter}

In this experiment, we evaluate how Path-Rank-Filter can select a subset of the adenomatous training images with the strongest adenomatous features for CycleGAN so that the generated images will also have a strong presence of features representing the desired class. 
For the three adenomatous classes of polyps (tubular, tubulovillous/villous, and sessile serrated), we apply CycleGAN using Path-Rank-Filter with filtration parameter values of  $\alpha=\{\frac{1}{2}, \frac{1}{4}, \frac{1}{8}, \frac{1}{16}, \frac{1}{32} \}$ on the 9054 normal colonic mucosa images in our training set to generate 9054 images of the target adenomatous class. 
We then measure the prominence of adenomatous features in our generated images by using a pre-trained classifier to evaluate the percent of generated images that were actually classified as the intended target class (Table \ref{tab:ablation}). 

\vspace{5mm}
\begin{table}[ht]
\setlength{\abovecaptionskip}{5pt}

\centering
\begin{tabular}{l  c c c c c c}
    \toprule
     Polyp Class & $\alpha = 1$ & $\alpha = \nicefrac{1}{2}$ & $\alpha = \nicefrac{1}{4}$ & $\alpha = \nicefrac{1}{8}$ & $\alpha = \nicefrac{1}{16}$ & $\alpha = \nicefrac{1}{32}$\\
     \midrule
     TA & 35.4 & 64.4 & 79.6 & 87.6 & 89.2 & \bf{93.8} \\
     TVA & 32.7 & 67.3 & 49.4 & 63.1 & 85.9 & \bf{86.1}\\
     SSA & 37.0 & 20.9 & 21.5 & 38.5 & 23.4 & 43.7\\
     \bottomrule
\end{tabular}
\caption{Percent of synthetic images generated by a CycleGAN with various $\alpha$ parameters for Path-Rank-Filter that were classified by a pre-trained classifier as the intended class. 9054 synthetic images were evaluated for each class and $\alpha$ value. TA: tubular adenoma, TVA: tubulovillous/villous adenoma, SSA: sessile serrated adenoma.}
\label{tab:ablation}
\end{table}

Based on this evaluation metric, Path-Rank-Filter substantially enhanced adenomatous features in generated images for TA and TVA. 
For these two classes, the highest classification performance was at $\alpha=\frac{1}{32}$, with the pre-trained classifier correctly detecting 93.8\% of generated images for TA and 86.1\% for TVA. 
These high accuracies seem to reflect the nature of these two adenomatous classes, for which images in the training set reflect a range of features.
TA images are defined by hyperchromatic, pencillate nuclei; pathologists will label both images with small hints of pencillate nuclei and obviously strong tubular features as tubular.
Of the same nature, TVA images are characterized by finger-like extensions with hyperchromatic, pencillate nuclei, and therefore some images will have more villous features than others.

For SSA, on the other hand, Path-Rank-Filter did not significantly improve the performance. 
We hypothesize that this result reflects the differing nature of SSAs, which are classified by the presence of a single broad-based crypt. 
Unlike TAs and TVAs, SSAs do not present a spectrum of histological features, and so it makes sense that Path-Rank-Filter does not choose a better subset of SSAs for training CycleGAN, and therefore generated images did not exhibit stronger features of SSAs. 

\begin{figure}[ht]
    \setlength{\abovecaptionskip}{5pt}
    \includegraphics[width=\linewidth]{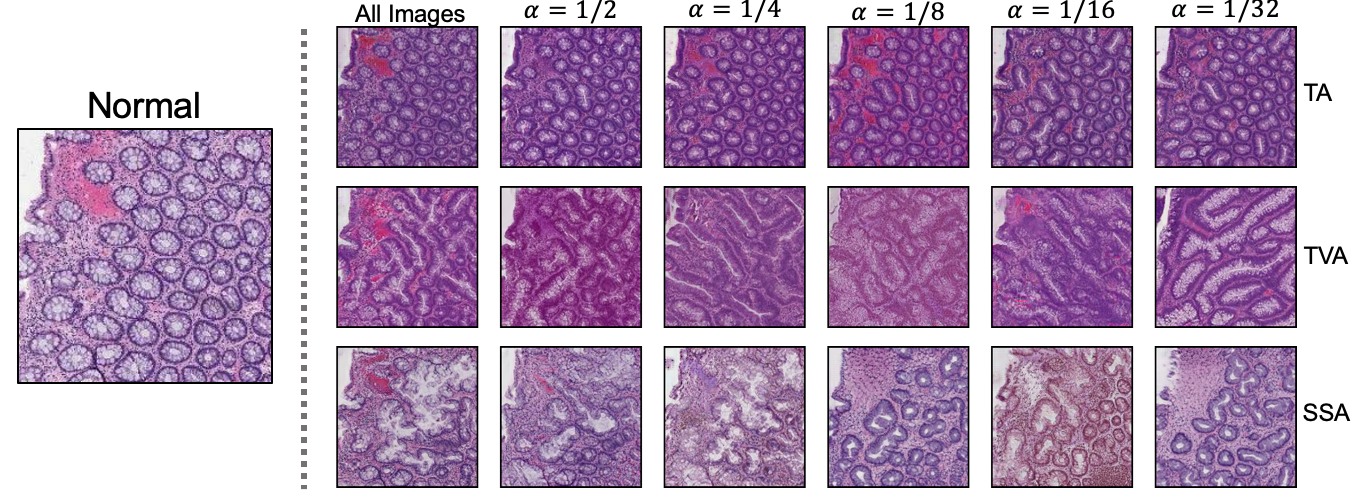}
    \caption{CycleGAN's generated images for different values of $\alpha$. For instance, $\alpha = 1/4$ means that the top 25\% of images with the highest output probabilities from a ResNet were used to train CycleGAN. TA: tubular adenoma, TVA: tubulovillous/villous adenoma, SSA: sessile serrated adenoma. For TA and TVA, adenomatous features were enhanced at smaller $\alpha$ values.}
    \label{fig:ablation}
\end{figure}

Furthermore, we select example images to examine the histologic features as we use different filtration parameters (Figure \ref{fig:ablation}). 
For TA, we see that CycleGAN transforms normal crypts by introducing pencillate nuclei into the crypt borders, altering cell color, and merging small crypts into more complex structures.
For TVA, crypts become more elongated and finger-like for smaller $\alpha$ parameters. 
For SSA, however, the quality of adenomatous features did not substantially improve with smaller $\alpha$ parameters; perhaps the SSA example shown when using all images for training $\alpha=1$ has the strongest features, although interpretations might differ among pathologists.
More examples of generated images for varying $\alpha$ are shown in Supplementary Figures \ref{fig:tubularablation} (TA), \ref{fig:villousablation} (TVA), and \ref{fig:sessileablation} (SSA).
Generated images of tubular adenomas after various epochs are shown in Supplementary Figure \ref{fig:tubularepochablation}.

\subsection{Evaluation by Pathologists}
\begin{figure}[ht]
    \setlength{\abovecaptionskip}{5pt}
    \includegraphics[width=\linewidth]{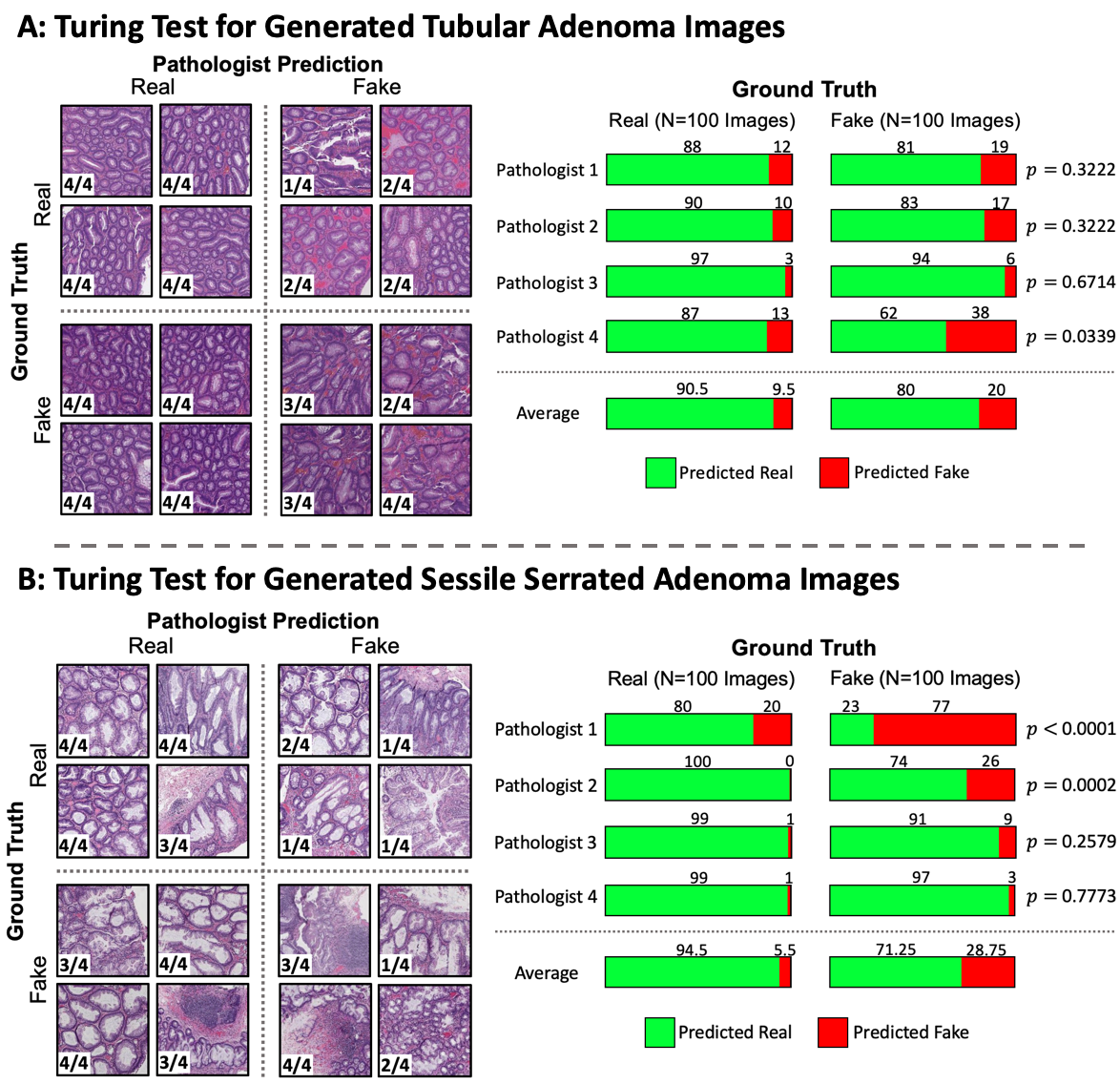}
    \caption{Results of Turing test for whether gastrointestinal pathologists could distinguish real and fake images of tubular adenomas (A) and sessile serrated adenomas (B). 
    Left: example real and generated images that were classified correctly and incorrectly by pathologists, with the number of pathologists who labeled the image as such denoted in the lower left corner.
    Right: evaluation of real and fake images by four pathologists. }
    \label{fig:turingimages}
\end{figure}

We further measure the quality of generated adenomatous images through clinical evaluation by four gastrointestinal pathologists. 
For the tubular and sessile serrated classes,\footnote{Manual evaluation is costly, and so we do not evaluate tubulovillous/villous adenoma in this paper.} the two least common classes in our dataset, we presented the four pathologists with a set of 200 unlabeled images: 100 real images and 100 generated (fake) images.
Each pathologist independently classified each image as either real or fake. 
As shown in Figure \ref{fig:turingimages}, at least half of the pathologists could not distinguish real and fake images at a statistically significant level, correctly distinguishing some fake images while also incorrectly labeling real images as fake.

We also perform statistical analysis on the pathologists' overall accuracies, using $x_0 = 0.5$ as the expected accuracy for random guessing and each pathologist's accuracy on the $n = 200$ images as $\hat{x}$ to calculate the $z$-score for each pathologist (Equation \ref{eq:zstat}). 
\begin{equation}
    z = \frac{\hat{x} - x_0}{\sqrt{\frac{x_0 (1 - x_0)}{n}}}
    \label{eq:zstat}
\end{equation}
We then calculate $p$ for each pathologist given the null hypothesis $H_0: \hat{x} = x_0$.
With this configuration, a $p$-value where $p < 0.05$ is statistically significant (i.e., the pathologist is able to distinguish between real and fake images).

For tubular adenoma images, only one pathologist was able to differentiate real images from synthetic images at a statistically significant level.
For sessile serrated adenoma images, two pathologists were able to distinguish between real and synthetic images at a statistically significant level.
Based on feedback from pathologists, fake sessile serrated adenoma images were easier to identify because our CycleGAN model created a subtle mosaic-like pattern in the whitespace of images.
Sessile serrated adenomas tended to have more whitespace because they are defined by a single large crypt (of mostly whitespace), which might explain why it was easier to detect fake sessile serrated adenomas than tubular adenomas. 


\subsection{Improving Classifier Performance}
\label{auc}

\begin{figure}[ht]
    \setlength{\abovecaptionskip}{5pt}
    \includegraphics[width=\linewidth]{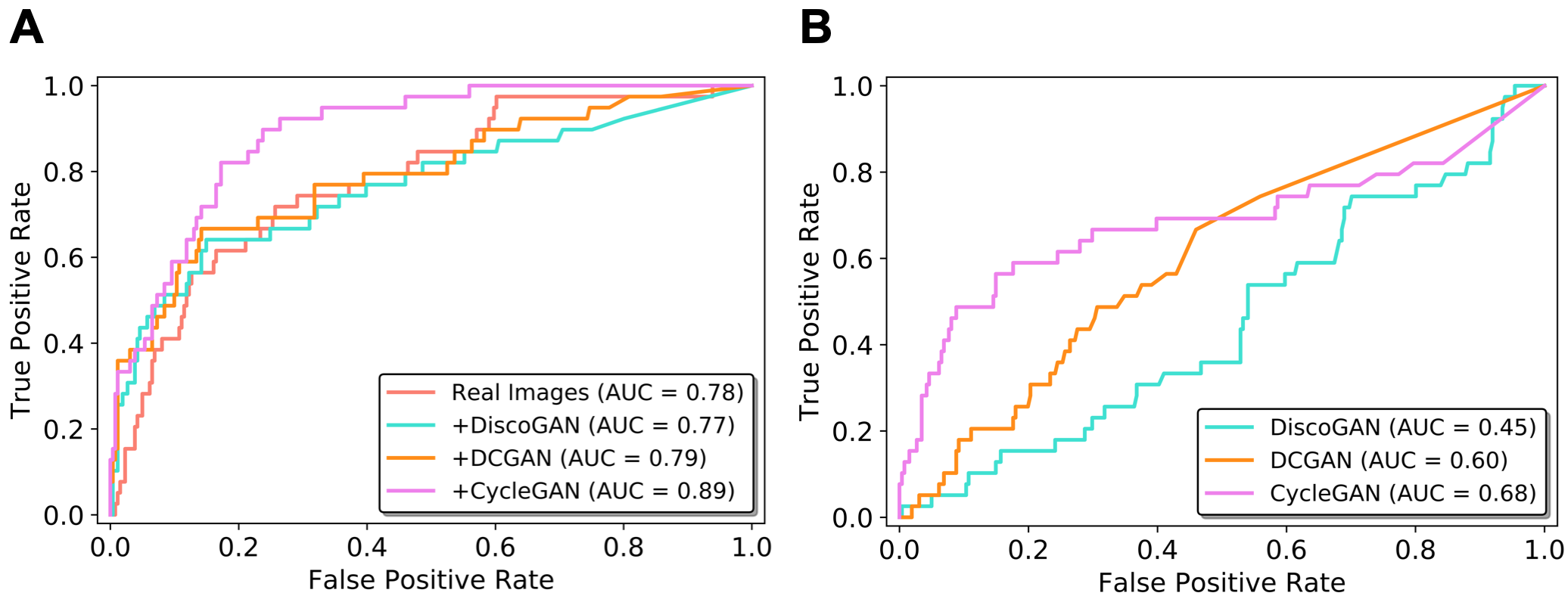}
    \caption{\textbf{A}: AUCs of ResNets trained on real images with synthetic images from different generative models given as additional training data. \textbf{B}: AUCs of ResNets trained without real images and with synthetic images from different generative models as the only available training data. In both experiments, the ResNet that was trained with CycleGAN's synthetic images had the highest AUC.}
    \label{fig:improveclassifier}
\end{figure}

Image translation can mitigate class imbalances in training sets by generating synthetic images of rare classes. 
We generated synthetic images of sessile serrated adenomas (only represented by 3\% of the training set) and used them as augmented data for training a ResNet classifier to distinguish between hyperplastic polyps (benign) and sessile serrated adenomas (adenomatous), a clinically important task in colorectal cancer screening \citep{Korbar2017b, Korbar2017}.  
We applied CycleGAN to all 9054 normal colonic mucosa images in our training set to generate 9054 images of the sessile serrated class, and added these images into the training set. 
Then, we used this dataset for training a ResNet and evaluated it on a test set of 261 hyperplastic polyp images and 39 sessile serrated adenoma images, comparing our ResNet's performance with that of ResNets trained on generated data from DiscoGAN and DCGAN, as well as ResNets trained without augmented data (Figure \ref{fig:improveclassifier}A).
Including CycleGAN-generated images for training boosted classification AUC by over 10\%, outperforming DCGAN-generated images and DiscoGAN-generated images.

We also train ResNet on a training set consisting of the same real hyperplastic images but with synthetic images as the only available sessile serrated adenoma images (Figure \ref{fig:improveclassifier}B). 
Once again, the model trained on CycleGAN-generated images outperformed the models trained on DCGAN-generated images and DiscoGAN-generated images by 8\% and 23\%, respectively. 
In both experiments, the ResNet that was trained using CycleGAN-generated images achieved the highest AUC.

\section{Limitations and Discussion}

Although we show some promising results in terms of image quality and ability to improve the performance of a ResNet classifier, our study has notable limitations. 
First, fair manual evaluation of images is non-trivial. 
Even though the pathologists in our study have years of experience examining colorectal polyp slides, these Turing tests do not perfectly reflect image quality, since pathologists do not distinguish real and fake data as a task in clinical practice. 
Furthermore, variation in results suggest that distinguishing fake images might depend highly on the individual pathologists, and some pathologists reported that they could better distinguish real and fake images as they saw more images. 
Finally, we only showed pathologists fixed-sized tiles of images; generating an entire high-resolution slide with high-quality features is a substantially more challenging task.

In terms of improving classifier training, we had hoped that training with synthetic data would achieve the same performance as training with real data, but a ResNet trained on only synthetic SSA images achieved an AUC of only 0.68 (Figure \ref{fig:improveclassifier}), much lower than the AUC of a classifier trained on both real and synthetic data (0.89). 
This result suggests that although the quality of a single generated image might be comparable to that of a single real image, the quality of the set of generative images likely does not match that of a set of real images.

Our paper has explored image translation for data augmentation in colorectal histopathology images. 
Whereas most work in generative data augmentation focuses on generating images from random noise, we note that images from other classes might be helpful in the field of histopathology and therefore take an image translation approach. 
Future work might include evaluating our method on other datasets to evaluate the generalizability of our approach.



\subsubsection*{Acknowledgments}
This research was supported in part by National Institute of Health grants R01LM012837 and P20GM104416.
We thank Naofumi Tomita for helpful feedback on the study design.

\newpage 
\bibliographystyle{acl_natbib}
\bibliography{bibliography}

\newpage

\renewcommand\thefigure{\arabic{figure}}
\setcounter{figure}{0}
\renewcommand{\figurename}{Supplementary Figure}

\setcounter{section}{0}

\section*{Supplementary Data}


\begin{figure}[ht]
    \setlength{\abovecaptionskip}{5pt}
    \includegraphics[width=\linewidth]{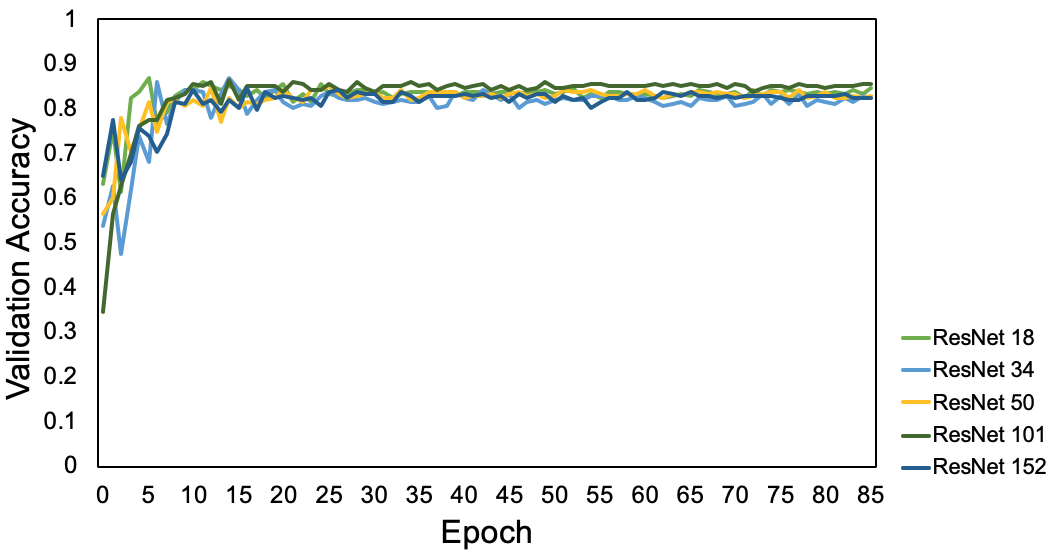}
    \caption{Validation accuracy of ResNet classifiers of varying depth. Performance did not improve substantially for deeper networks.}
    \label{fig:resnetlayers}
\end{figure}

\begin{figure}[ht]
    \setlength{\abovecaptionskip}{5pt}
    \includegraphics[width=\linewidth]{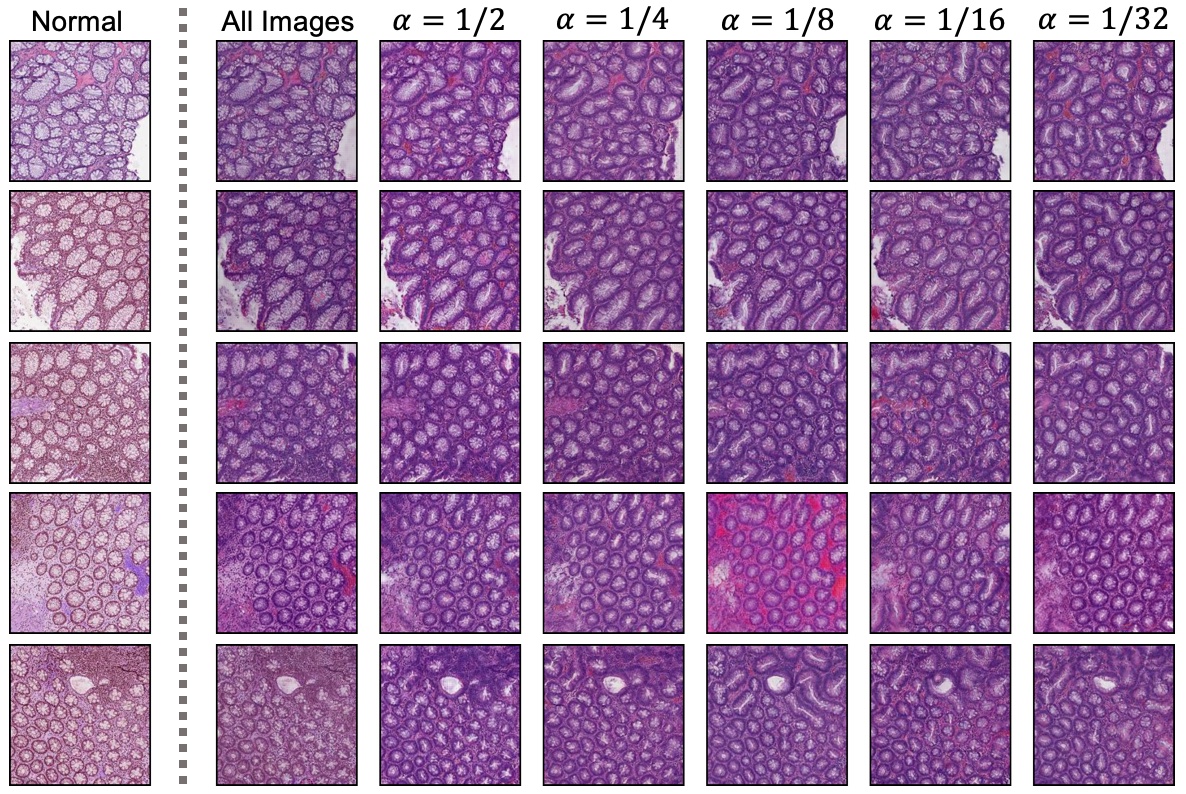}
    \caption{Examples of tubular adenoma images generated by CycleGAN with Path-Rank-Filter at varying $\alpha$ levels. Adenomatous features were enhanced at lower $\alpha$.}
    \label{fig:tubularablation}
\end{figure}

\begin{figure}[ht]
    \setlength{\abovecaptionskip}{5pt}
    \includegraphics[width=\linewidth]{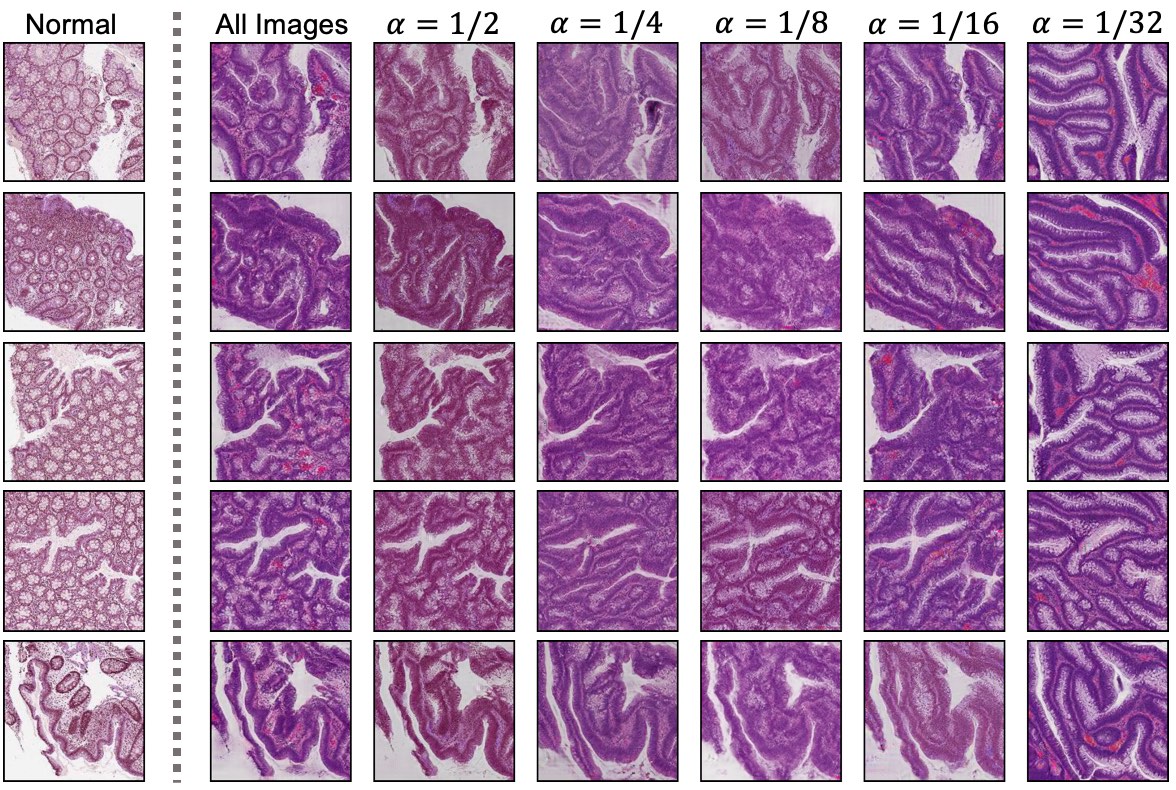}
    \caption{Examples of tubullovillous/villous adenoma images generated by CycleGAN with Path-Rank-Filter at varying $\alpha$ levels. Adenomatous features were enhanced at lower $\alpha$.}
    \label{fig:villousablation}
\end{figure}

\begin{figure}[ht]
    \setlength{\abovecaptionskip}{5pt}
    \includegraphics[width=\linewidth]{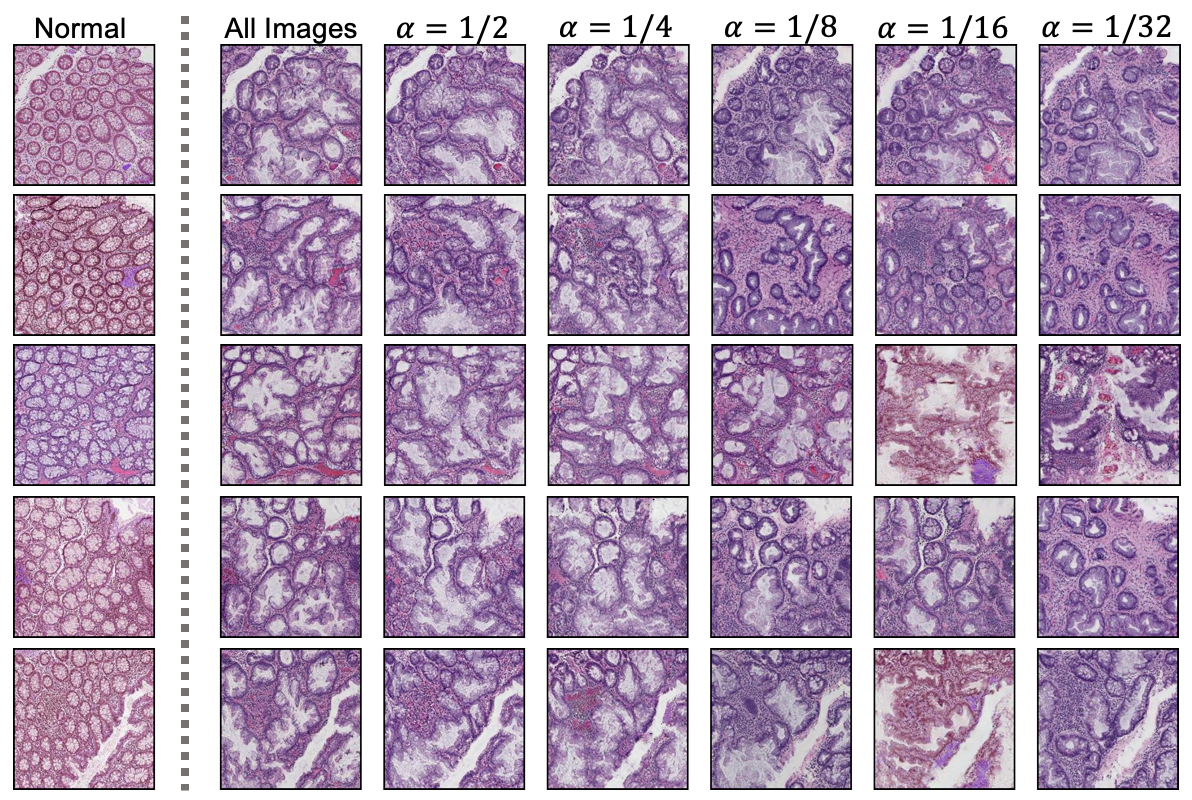}
    \caption{Examples of sessile serrated adenoma images generated by CycleGAN with Path-Rank-Filter at varying $\alpha$ levels. Using lower $\alpha$ values did not enhance the features of seesile serrated adenomas.}
    \label{fig:sessileablation}
\end{figure}

\begin{figure}[ht]
    \setlength{\abovecaptionskip}{5pt}
    \includegraphics[width=\linewidth]{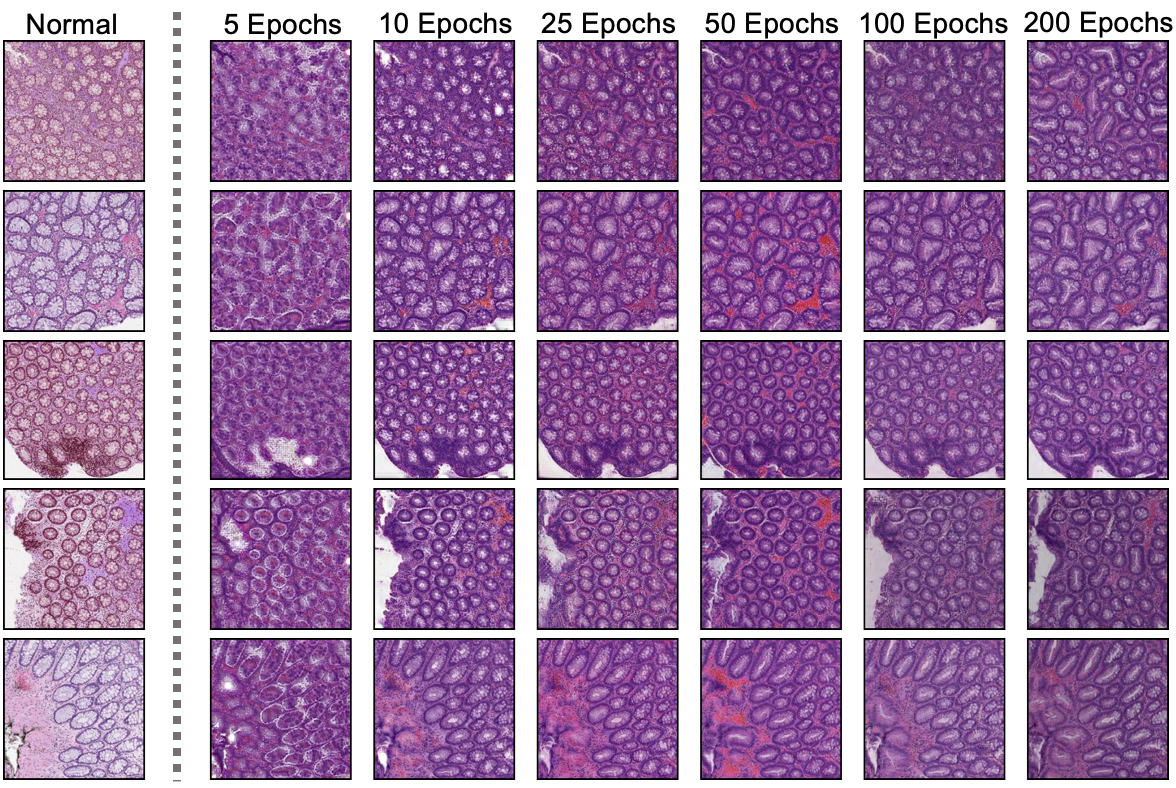}
    \caption{Examples of tubular adenoma images generated by CycleGANs trained for 5, 10, 25, 50, 100, and 200 epochs. Convergence occurred at approximately 200 epochs.}
    \label{fig:tubularepochablation}
\end{figure}

\begin{figure}[ht]
    \setlength{\abovecaptionskip}{5pt}
    \includegraphics[width=\linewidth]{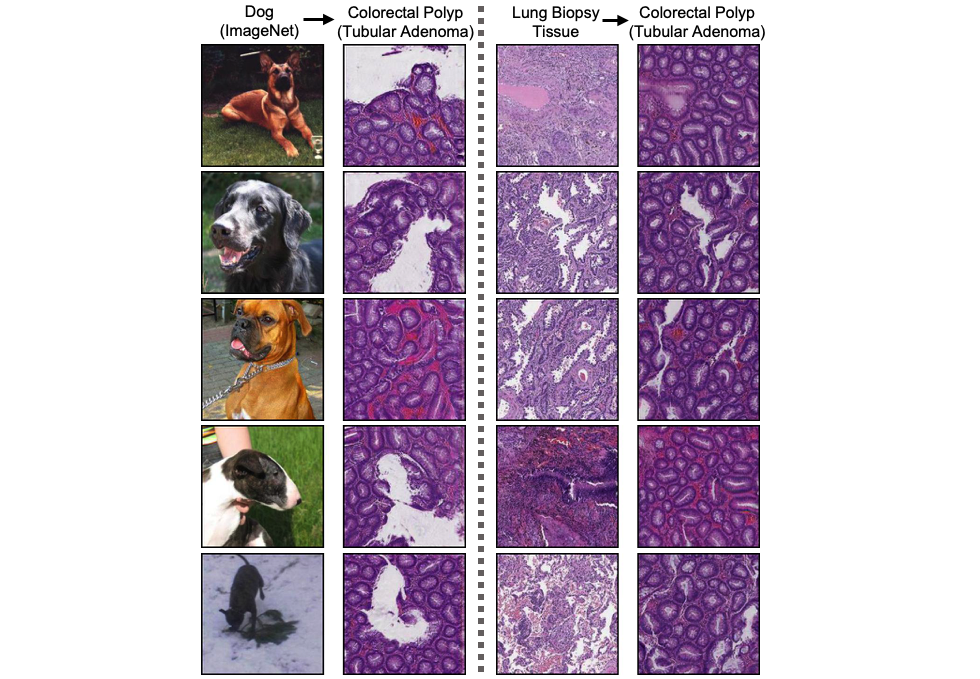}
    \caption{Tubular adenoma images generated with CycleGAN using dogs from ImageNet and lung biopsy tissue samples as source domains.}
    \label{fig:transformationbase}
\end{figure}

\end{document}